# Moment of inertia of a trapped superfluid gas of atomic Fermions

M. Farine[1], P. Schuck[2], X. Viñas[3]

[1]Ecole Navale, Lanvéoc-Poulmic, 29240 Brest-Naval, France
[2]Institut des Sciences Nucléaires, Université Joseph Fourier, CNRS-IN2P3
53, Avenue des Martyrs, F-38026 Grenoble Cedex, France
[3]Departament d' Estructura i Constituents de la Mat'eria Facultat de F'isica,
Universitat de Barcelona Diagonal 647, E-08028 Barcelona, Spain

**Abstract:** The moment of inertia $\Theta$ of a trapped superfluid gas of atomic Fermions ($^6$Li) is calculated as a function of two system parameters: temperature and deformation of the trap. For moderate deformations at zero temperature the moment of inertia takes on the irrotational flow value. Only for T very close to $T_c$ rigid rotation is attained. For very strong trap deformations the moment of inertia approaches its rigid body value even in the superfluid state. It is proposed that future measurements of the rotational energy will unambiguously reveal whether the system is in a superfluid state or not.

## 1. Introduction

The advent in 1995 of Bose-Einstein-Condensation (BEC) of atomic Bosons in magnetic traps certainly represents a milestone in the study of bosonic many body quantum systems. This is so because a systematic study of these systems, starting with the free particle case, as a function of increasing density, particle number, and other system parameters seems possible and has largely already been cut into and is progressing with a rapid pace [1,2]. On the other hand the recent experimental achievement of trapping $^6$Li atoms and other fermionic alkali atoms [3] also spurs the hope that for the fermionic many body problem, as much progress will be made in the near future as for the bosonic systems. For instance the fact that trapped spin-polarized $^6$Li atoms which feel a strong attraction in the triplet channel (scattering length: a = -2063 $a_0$ with $a_0$ the Bohr radius) may undergo a phase transition to the superfluid state has recently provoked a number of theoretical investigations [4]. One major question which is under debate is how to detect the superfluidity of such a fermionic system, since in contrast to a bosonic system the density of a Fermionic system is hardly affected by the transition to the superfluid state [5]. Several proposals such as the study of the decay rate of the gas or of the scattering of atoms off the gas have been advanced [4]. Though such investigations may give precious indications of a possible superfluid phase, we think that in analogy with nuclear physics, a measurement of the moment of inertia certainly would establish an unambiguous signature of superfluidity. To measure the spin and the rotational energy of trapped atoms definitely is a great challenge for the future. However, in nuclear physics, where γ-spectroscopy is extremely well developed, the strong reduction of the moment of inertia with respect to its rigid body value has been considered as a firm indicator of nucleon superfluidity immediately after the discovery of nuclear rotational states almost half a century back [5]. Therefore awaiting future experimental achievement also for trapped fermionic atoms, it is our intention in this work to give some theoretical estimates of the moment of inertia as a function of some system parameters such as deformation of the traps or temperature of the gas. In this study we can largely profit from the experience nuclear physicists have accumulated over the last decades in describing such phenomena. The expectation is indeed that there will be a great analogy between the physics of confined atomic Fermions and what one calls in nuclear physics the liquid drop part of the nucleus. As astonishing as it may seem assemblies of fermions containing no more than ~200 particles (nucleons) already exhibit an underlying macroscopic structure well known from the Bethe-Weizsaecker formula for nuclear masses [5]. In superfluid rotating nuclei as early as 1959 Migdal proposed a statistical description of the nuclear moment of inertia [6] which grasped the essential physics of a self contained rotating superfluid Fermi liquid drop and which serves as a reference even today.

In the present work we will cast Migdal's approach into the more systematic language of the Thomas Fermi theory which together with its extensions is applied extensively since decades to normal fluid but also to superfluid nuclei [5,7,8]. It is fortunate that we can profit from this experience for the description of trapped fermions, since their number of order $10^5$, together with the smoothness of the potential, certainly turns a statistical description into a very precise tool. On the other hand it may not be excluded that in the future the study of much smaller systems of trapped atomic Fermions with numbers ~$10^2$ may be studied probably revealing many analogies with real nuclei such as shell structure etc. The investigation of the transition from microscopic to macroscopic as the number of particles is increased continuously may then become a very interesting field also in the case of atomic Fermions. In detail our paper is organized as follow. In sections 2 and 3 we review the Thomas Fermi approach to inhomogeneous superfluid Fermi systems. In section 4 first the so-called Inglis part of the moment of inertia of a rotating superfluid and confined gas of atomic Fermions is presented. Second the influence of the reaction of the pair field on the moment of inertia is calculated. It is shown that this leads to the irrotational flow value in the limit of strong pairing. In section 5 the current distributions in the superfluid and normal fluid regimes are contrasted. In section 6 the numerical results are presented in detail. Finally in section 7 we discuss our results and present some conclusions.

## 2. Thomas Fermi approach to fermionic atoms in deformed traps

As in the boson case the Thomas Fermi (TF) approach [5] to trapped atomic gases is rendered extremely simple by two facts: the smoothness of the traps (harmonic oscillator) and the large interparticle distance which makes a pseudopotential approximation to atomic interactions valid. Let us therefore write down the TF equation for a double spin polarized system of trapped ($^6$Li) atoms in the normal fluid state. For convenience we first consider the system at zero temperature T discussing the T $\neq$ 0 case later on. In TF approximation the distribution function is given by,

$$f(\mathbf{R},\mathbf{p}) = \theta(\mu - H_{cl}) \tag{2.1}$$

with,

$$H_{cl} = \frac{p^2}{2m} + V_{ex}(\mathbf{R}) - g\rho(\mathbf{R}) \quad (2.2)$$

Here $\mu$ is the chemical potential, and $V_{ex}(\mathbf{R})$ the trap potential supposed to be of harmonic form. The density $\rho(\mathbf{R})$ is obtained from the self-consistency equation,

$$\rho(\mathbf{R}) = \int \frac{d^3 p}{(2\pi\hbar)^3} f(\mathbf{R},\mathbf{p}) = \frac{1}{6\pi^2} k_F^3(\mathbf{R}) \quad (2.3)$$

with,

$$k_F(\mathbf{R}) = \sqrt{\frac{2m}{\hbar^2}(\mu - V_{ex}(\mathbf{R}) + g\rho(\mathbf{R}))} \quad (2.4)$$

the local Fermi momentum. The coupling constant g is related to the scattering length in the same way as in the Bose gases [1,2] via:

$$g = \frac{4\pi\hbar^2 |a|}{m} \quad (2.5)$$

The TF equation (2.3) leads to a cubic equation for the self-consistent density, which can be solved without problem as a function of the external potential. In this paper our main interest will be the study of the moment of inertia of a rotating condensate. Since the study is very much simplified assuming that the self consistent potential is again a harmonic oscillator and since the effect of the attractive interaction between the atoms essentially results in a narrowing of the self consistent potential with respect to the external one we will use instead of the exact TF solution for the density the following trial ansatz for the local Fermi momentum:

$$k_F^{trial}(\mathbf{R}) = \sqrt{\frac{2m}{\hbar^2}\left(\mu - \frac{m}{2}\left(\omega_\perp^2(R_x^2 + R_y^2) + \omega_z^2 R_z^2\right)\right)} \qquad (2.6)$$

where $\omega_\perp$ and $\omega_z$ are the variational parameters. The chemical potential is determined from the particle number condition,

$$N = \int d^3r \rho^{trial}(r) \qquad (2.7)$$

and the kinetic energy density is given by,

$$\tau(\mathbf{R}) = \int \frac{d^3p}{(2\pi\hbar)^3} \frac{p^2}{2m} f(\mathbf{R},\mathbf{p}) = \frac{1}{10\pi^2}\left[k_F^{trial}(\mathbf{R})\right]^5 \qquad (2.8)$$

We then can analytically calculate the total energy,

$$E(\omega_\perp, \omega_z) = \int d^3R \left[\tau(\mathbf{R}) + V_{ex}(\mathbf{R})\rho(\mathbf{R}) - \frac{g}{2}\rho^2(\mathbf{R})\right] \qquad (2.9)$$

as a function of $\omega_\perp$ and $\omega_z$. Minimizing this expression with respect to $\omega_\perp$ and $\omega_z$ for a given external deformed harmonic oscillator potential,

$$V_{ex} = \frac{m}{2}\left(\omega_{0\perp}^2(R_x^2 + R_y^2) + \omega_{0z}^2 R_z^2\right) \qquad (2.10)$$

leads to the variational solution. For the spherical case $\omega_\perp = \omega_z = \omega$, this is shown in Fig. 1. We see that this approximation to the TF equation is quite reasonable. For $\omega_0 = 6.9\,\text{s}^{-1}$, the value of the variational frequency is $\omega = 7.69\,\text{s}^{-1}$ that is $\omega > \omega_0$, implying a compression of the density. Increasing $\omega$ by 6% from its variational value allows an almost perfect reproduction of the full TF solution. We will adopt this latter value in all our forthcoming calculations. In order to obtain the deformed case we simply squeeze the potential using a deformation depending on one parameter,

$$\delta = \frac{\omega_z}{\omega_\perp} \tag{2.11}$$

Frequencies $\omega_\perp$ and $\omega_z$ are then defined as functions of the deformation parameter by,

$$\omega_\perp = \omega \delta^{-(1/3)}$$
$$\omega_z = \omega \delta^{(2/3)} \tag{2.12}$$

in order to keep constant for any value of $\delta$ the volume,

$$(\omega_\perp)^2 \omega_z = \omega^3. \tag{2.13}$$

From now on we therefore will use for the non superfluid Wigner function at zero temperature the expression,

$$f(\mathbf{R},\mathbf{p}) = \theta\left(\mu - \frac{p^2}{2m} - \frac{m}{2}\left(\omega_\perp^2(R_x^2 + R_y^2) + \omega_z^2 R_z^2\right)\right) \tag{2.14}$$

with $\omega_z, \omega_\perp$ from (2.12) and $\mu$ determined from the particle number condition.

## 3. The superfluid case

Since trapped spin polarized $^6$Li atoms feel a strong attractive interaction in the triplet channel the system very likely will undergo a transition to the superfluid state at some critical temperature $T_c$ as was discussed in detail in ref. [4]. As we have pointed out in the introduction the superfluid state will unambiguously reveal itself in its value of the moment of inertia. At the moment the measurement of angular momenta of trapped Bose or Fermi gases has not been achieved and represents a great future challenge to the experimenters. In order to establish how the two essential system parameters which are the value of the gap, i.e. the temperature and the deformation of the external trap influence the value of the moment of inertia, we will now proceed to its evaluation in the superfluid state.

Since we are dealing with an inhomogeneous system, even in the non rotating case the gap is actually a non local quantity $\Delta(\mathbf{r},\mathbf{r}')$ or in Wigner space $\Delta(\mathbf{R},\mathbf{p})$. It has been shown in [8,5] that to lowest order in $\hbar$ the gap equation is given by,

$$\Delta(\mathbf{R},\mathbf{p}) = \int \frac{d^3k}{(2\pi\hbar)^3} v(\mathbf{p}-\mathbf{k}) \frac{\Delta(\mathbf{R},\mathbf{k})}{2E(\mathbf{R},\mathbf{k})} \tanh\left(\frac{E(\mathbf{R},\mathbf{k})}{2T}\right) \qquad (3.1)$$

where $E(\mathbf{R},\mathbf{p})$ is the quasiparticle energy,

$$E(\mathbf{R},\mathbf{p}) = \sqrt{\left[\frac{p^2 - p_F^2(\mathbf{R})}{2m^*(\mathbf{R})}\right]^2 + \Delta^2(\mathbf{R},\mathbf{k})} \qquad (3.2)$$

with $p_F(\mathbf{R}) = \hbar k_F(\mathbf{R})$ the local Fermi momentum (2.4). Since the effective mass $m^*$ is so far unknown for the trapped gases of atomic Fermions we will take $m^* = m$. Furthermore, for the time being, as in [4], we will eliminate the interatomic potential v in (3.1), expressing it by the scattering length (2.5). We then obtain [4],

$$\Delta(\mathbf{R},\mathbf{p}) = g \int \frac{d^3k}{(2\pi\hbar)^3} \left[\frac{\tanh\left(\frac{E(\mathbf{R},\mathbf{k})}{2T}\right)}{2E(\mathbf{R},\mathbf{k})} - \frac{P}{2(\varepsilon_k - \varepsilon_F(\mathbf{R}))}\right] \Delta(\mathbf{R},\mathbf{k}) \qquad (3.3)$$

where P stands for principal value, $\varepsilon_k = \frac{\hbar^2 k^2}{2m}$ and $\varepsilon_F = \frac{\hbar^2 k_F^2}{2m}$. At zero temperature, as described in [9], (3.3) can be solved analytically in the limit $\frac{\Delta(\mathbf{R}, p_F(\mathbf{R}))}{\varepsilon_F(\mathbf{R})} \to 0$. The result is given by,

$$\Delta_F(\mathbf{R}) \equiv \Delta(\mathbf{R}, k_F(\mathbf{R})) = 8e^{-2}\varepsilon_F(\mathbf{R}) e^{-\frac{\pi}{2k_F(\mathbf{R})|a|}} \tag{3.4}$$

A posteriori one can verify that $\frac{\Delta_F}{\varepsilon_F} \ll 1$ for all values of R and therefore (3.4) is an excellent approximation to (3.3). This also has been verified in [4]. For 2.865 x $10^5$ $^6$Li atoms, the case considered in [4], the gap is shown for a spherical trap as a function of the radius in Fig. 2.

For the determination of the critical temperature $T_c$ and, later on, for the moment of inertia we will need the value $\Delta$ of the gap averaged over the states $|n\rangle$ at the Fermi level,

$$\Delta(\varepsilon_F) \equiv \Delta = Tr(\hat{\Delta}\hat{\rho}(\varepsilon_F)) \tag{3.5}$$

with,

$$\hat{\rho}(\varepsilon_F) = \frac{1}{g(\varepsilon_F)}\sum_n |n\rangle\langle n|\delta(\varepsilon_F - \varepsilon_n) = \frac{1}{g(\varepsilon_F)}\delta(\varepsilon_F - H) \tag{3.6}$$

where $|n\rangle$ and $\varepsilon_n$ are the shell model states and energies and,

$$g(\varepsilon_F) = \sum_n \delta(\varepsilon_F - \varepsilon_n) = Tr\delta(\varepsilon_F - H) \tag{3.7}$$

is the level density at the Fermi energy.

It has been shown in [10] that again the TF approximation leads to an excellent average value,

$$\Delta = \frac{1}{g^{TF}(\varepsilon_F)} \int \frac{d^3R\, d^3p}{(2\pi\hbar)^3} \Delta_F(R)\delta(\varepsilon_F - H_{cl}) \tag{3.8}$$

In the spherical case with $\Delta_F(R)$ from (3.4) all integrals but the radial one can be performed analytically, the latter being done numerically. For the case shown in Fig. 2 one obtains,

$$\Delta = 16.4 \text{ nK} \tag{3.9}$$

One now can discuss the question whether it is the value (3.9) which determines the critical temperature $T_c$ or whether it is $\Delta_F(R=0) \approx 65\,\text{nK}$ as determined from Fig. 2. In the strict quantum mechanical sense the BCS equations should be solved in the shell model or better HF basis and then $T_c$ is a global parameter which must be determined from the solution of the quantum mechanical gap equation. Since we believe that the value in (3.9) comes rather close to the quantum mechanical value of the gap at the Fermi energy we think that (3.9) should determine $T_c$ which we can obtain from the usual BCS weak coupling relation [11] $\Delta = 1.76 T_c$ to be

$$T_c \approx 10\,\text{nK}. \tag{3.10}$$

On the other hand the temperature dependent BCS approach is certainly not an exact theory for finite systems and it may be that e.g. fluctuations and correlations break off the quantum coherence of the global single particle wave function and more local cluster structures are formed. In such cases it could be that superfluidity persists locally to higher temperature than the one deduced from (3.9). A precise answer to this question is certainly quite difficult and in view of this uncertainty we will later for the moment of inertia take the critical temperature as a parameter which can vary within certain limits determined by the gap parameter (3.9) or $\Delta_F(0)$ respectively. Independent of the question of the precise value of $T_c$ we will have to know the detailed T-dependence of the gap $\Delta(T)$ which however, in BCS theory, given $\Delta_0$ and $T_c$, is determined by the universal function $\frac{\Delta(T)}{\Delta(0)}$ in terms of $\frac{T}{T_c}$. This function is determined from the solution of the equation [11],

$$-\ln\left(\frac{\Delta(T)}{\Delta(0)}\right) = A\left(\frac{\Delta(T)}{T}\right)$$

with, (3.11)

$$A(u) = \int_0^\infty dy \frac{1}{\sqrt{y^2+u^2}}\left(1-\tanh\left(\frac{\sqrt{y^2+u^2}}{2}\right)\right)$$

For completeness it is again shown in fig. 3. This T-dependence of the gap we will later use for the evaluation of the moment of inertia.

## 4. Moment of inertia

The moment of inertia of a rotating nucleus has fully been formulated in linear response theory (i.e. RPA) by Thouless and Valatin [12]. The corresponding expression is therefore called, in the nuclear physics literature [5], the Thouless-Valatin moment of inertia. It consists of two parts, the so-called Inglis term, which describes the free gas response, and the part, which accounts for the reaction of the mean field and pair potential to the rotation. In the superfluid case the Inglis part has been generalized by Belyaev [13] and the linear reaction of the gap parameter onto the value of the moment of inertia was first evaluated, together with the Inglis term, by Migdal [6]. The reaction of the HF field on the rotation is a minor effect and we will neglect it in this work. We therefore will write the moment of inertia as a sum of the Inglis-Belyaev term $\Theta_{I-B}$ and the Migdal term $\Theta_M$. In total,

$$\Theta = \Theta_{I-B} + \Theta_M \tag{4.1}$$

In order to derive an expression for $\Theta$ in linear response theory we will use the Gorkov approach described in detail in many text books (in what follows we will use the notation of [14]). Since in addition the derivation of the linear response for $\Theta$ is given in the original article of Migdal [6] and rerepresented in a more elaborate version in [7], we will be very short here and only give more details where in our opinion the presentations in [6,7 ] may not be entirely explicit. Let us start writing down the Gorkov equations in matrix notation,

$$\left(-\frac{\partial}{\partial \tau} - H + \mu\right) G = 1 - \Delta F^+$$

$$\left(\frac{\partial}{\partial \tau} + H^* + \mu\right) F^+ = \Delta^* G$$

(4.2)

with

$$H = H_0 - \Omega L_x \equiv H_0 + H_1 \tag{4.3}$$

where $H_0$ is the shell model Hamiltonian (2.2 ) and

$$L_x = r_y p_z - r_z p_y$$

the angular momentum operator corresponding to a rotation with angular frequency $\Omega$ around the x-axis. In (4.2) G and F are the normal and anomal Matsubara Green's functions (see Chapt. 51 of [ 14]),

$$G_{nn'} = -\langle T_\tau a_n(\tau) a_{n'}^+(\tau')\rangle$$

$$F^+{}_{nn'} = -\langle T_\tau a_n^+(\tau) a_{n'}^+(\tau')\rangle$$

(4.4)

Linearising (4.2) with respect to $H_1$, that is $G = G_0 + G_1$, $F^+ = F_0^+ + F_1^+$ and $\Delta = \Delta_0 + \Delta_1$ (as mentioned we will neglect the influence of the rotational field on $H_0$) one obtains for (4.2),

$$G_1 = G_{1I-B} + G_{1M} \tag{4.5}$$

with,

$$G_{1I-B} = G_0 H_1 G_0 + F_0^+ H_1^* F_0^+ \tag{4.6}$$

$$G_{1M} = -G_0 \Delta_1 F_0^+ - F_0^+ \Delta_1^* G_0$$

and,

$$F_1^+ = D_0 H_1^* F_0^+ + F_0^+ H_1 G_0 - F_0^+ \Delta_1 F_0^+ + D_0 \Delta_1^* G_0 \tag{4.7}$$

where,

$$D = \frac{i\omega_n - H_0}{\omega_n^2 + H_0^2 + \Delta_0^2} \quad ; \quad G_0 = \frac{i\omega_n + H_0}{\omega_n^2 + H_0^2 + \Delta_0^2} \quad ; \quad F_0^+ = \frac{\Delta_0}{\omega_n^2 + H_0^2 + \Delta_0^2}$$

and $\omega_n$ are the Matsubara frequencies [14].

In (4.5,4.6) we have split the first order Green's function in an obvious notation into the Inglis-Belyaev and Migdal contributions. For the latter one needs the linear reaction of the pair field to the rotation. We later will see how this can be determined from (4.7). First let us, however, evaluate the I-B part of the moment of inertia.

### 4.1 The Inglis-Belyaev part of the moment of inertia

The I-B part of the moment of inertia can be evaluated without the knowledge of $\Delta_1$ i.e. without the use of (4.7). The density response corresponding to $G_{1I-B}$ of (4.5) is evaluated from the limit $\tau' \to \tau^+$ or from summing over the Matsubara frequencies in the upper half plane (see Ch. 7 of [14]). One obtains the well known result [5, 6, 7, 14],

$$(\rho_{1I-B})_{nn'} = \langle n|L_x|n'\rangle F_{nn'} \tag{4.8}$$

with,

$$F = F_+(1 - f - f') + F_-(f - f') \tag{4.9}$$

where,

$$F_\pm = F_\pm(\varepsilon_n, \varepsilon_{n'}) = \frac{E_n E_{n'} \mp \xi_n \xi_{n'} - \Delta(\varepsilon_n)\Delta(\varepsilon_{n'})}{2 E_n E_{n'} (E_n \pm E_{n'})} \tag{4.10}$$

$$f = f(\varepsilon_n) = \frac{1}{1 + e^{E_n/T}} \quad ; \quad f' = f(\varepsilon_{n'}) \tag{4.11}$$

and,

$$E_n = \sqrt{\xi_n^2 + \Delta^2(\varepsilon_n)} \quad ; \quad \xi_n = \varepsilon_n - \mu \tag{4.12}$$

are the quasiparticle energies with as ingredient $\varepsilon_n$, the shell model energies. The gap parameters $\Delta_n$ have been replaced in (4.10), in analogy to (3.5), to statistical accuracy by $\Delta(\varepsilon_n)$, the ones averaged over the energy shell. The moment of inertia is given by,

$$\Theta_{I-B} = Tr(L_x \rho_{1I-B}). \tag{4.13}$$

Since we are interested at temperatures $T \leq T_c$, which are very low with respect to the Fermi energy, we checked that one can to very good accuracy neglect in (4.9) the thermal factors (4.11). The only important temperature dependence of the moment of inertia therefore exists via the T-dependence of the gap. We thus will henceforth treat all formulas as in the T=0 limit keeping, however, the T-dependence of the gap. With this in mind we can write for the moment of inertia,

$$\Theta_{I-B} = \sum_{nn'} \iint d\omega d\omega' \delta(\omega - \varepsilon_n)\delta(\omega' - \varepsilon_{n'}) |\langle n|L_x|n'\rangle|^2 F_+(\omega, \omega') \qquad (4.14)$$

In this formula the important quantity to calculate to statistical accuracy is,

$$L_x^2(n,n') \equiv |\langle n|L_x|n'\rangle|^2 = Tr[(L_x)(|n'\rangle\langle n'|L_x|n\rangle\langle n|)]$$
$$= \int \frac{d^3R d^3p}{(2\pi\hbar)^3} (L_x)_W \left(|n'\rangle\langle n'|L_x|n\rangle\langle n|\right)_W \qquad (4.15)$$

where $O_W \equiv O(\mathbf{R}, \mathbf{p})$ means the Wigner transform of the operator $O$ [5]. To this purpose we again replace the density matrices $|n\rangle\langle n|$ and $|n'\rangle\langle n'|$ by their average on the energy shell (3.6),

$$|n\rangle\langle n| \to \hat{\rho}(\varepsilon_n)$$

We therefore obtain,

$$\Theta_{I-B} = \iint d\omega d\omega' \int \frac{d^3R d^3p}{(2\pi\hbar)^3} [(L_x)_W (L_x(\omega,\omega'))_W] F_+(\omega,\omega') \qquad (4.16)$$

with,

$$(L_x(\omega,\omega'))_W = [\delta(\omega' - \hat{H}_0)\hat{L}_x \delta(\omega - \hat{H}_0)]_W \qquad (4.17)$$

Introducing into (4.17) the Fourier representations of the two $\delta$-functions and transforming to center of mass and relative coordinates one obtains,

$$(L_x(\omega,\omega'))_W = \iint \frac{dTd\tau}{(2\pi\hbar)^2} e^{2iET} e^{i\varepsilon\tau/2} \left[e^{-iH_0 T} L_x\left(\frac{\tau}{2}\right) e^{-iH_0 T}\right]_W \qquad (4.18)$$

with,

$$E = \frac{\omega + \omega'}{2} \quad ; \quad \varepsilon = \omega - \omega' \qquad (4.19)$$

and,

$$O(t) = e^{iH_0 t} O(0) e^{-iH_0 t} \tag{4.20}$$

To lowest order in $\hbar$ we replace the triple operator product in (4.18) by the product of their Wigner transforms [5],

$$\lim_{\hbar \to 0}\left[ e^{-iH_0 T} L_x\left(\frac{\tau}{2}\right) e^{-iH_0 T} \right]_W = e^{-i2H_{0cl}T} L_x^{cl}\left(\frac{\tau}{2}\right) \tag{4.21}$$

and therefore,

$$(L_x(\omega,\omega'))_W = L_x(E,\varepsilon,\mathbf{R},\mathbf{p}) = \frac{1}{2}\delta(E - H_{0cl})\int \frac{d\tau}{(2\pi\hbar)} e^{\frac{i}{\hbar}\varepsilon\frac{\tau}{2}} L_x^{cl}\left(\frac{\tau}{2}\right) \tag{4.22}$$

with,

$$L_x^{cl}\left(\frac{\tau}{2}\right) = R_y\left(\frac{\tau}{2}\right) p_z\left(\frac{\tau}{2}\right) - R_z\left(\frac{\tau}{2}\right) p_y\left(\frac{\tau}{2}\right) \tag{4.23}$$

At this point the choice of our approximate self consistent potential of harmonic oscillator form (see 2.14) turns out to be very helpful, since the classical trajectories in (4.23) can be given analytically.

$$R_i\left(\frac{\tau}{2}\right) = R_i \cos\left(\hbar\omega_i \frac{\tau}{2}\right) + \frac{p_i}{m\omega_i}\cos\left(\hbar\omega_i \frac{\tau}{2}\right)$$

$$p_i\left(\frac{\tau}{2}\right) = p_i \cos\left(\hbar\omega_i \frac{\tau}{2}\right) + m\omega_i R_i \cos\left(\hbar\omega_i \frac{\tau}{2}\right) \tag{4.24}$$

with $i = x, y, z$ and $\omega_x = \omega_y = \omega_\perp$

In the phase space integral of (4.15), for reasons of symmetry, only the diagonal terms of $L_x^{cl}.L_x^{cl}(\tau/2)$ survive and therefore we obtain,

$$\int \frac{d^3R d^3p}{(2\pi\hbar)^3} L_x^{cl} L_x^{cl}\left(\frac{\tau}{2}\right) = \int d^3R \rho^{TF}(\mathbf{R})(R_y^2 + R_z^2)\cos\left(\hbar\omega_\perp \frac{\tau}{2}\right)\cos\left(\hbar\omega_z \frac{\tau}{2}\right)$$
$$+ \left(R_y^2 \frac{\omega_\perp}{\omega_z} + R_z^2 \frac{\omega_z}{\omega_\perp}\right)\sin\left(\hbar\omega_\perp \frac{\tau}{2}\right)\sin\left(\hbar\omega_z \frac{\tau}{2}\right) \tag{4.25}$$

where,

$$\rho^{TF} = \frac{1}{6\pi^2} \left[ \frac{2m}{\hbar^2} (E-V) \right]^{3/2} \qquad (4.26)$$

is the density in TF approximation.

The product of cosine and sine in (4.25) can be expressed in terms of the cosine of the sum and difference of the arguments and then the $\tau$-integral in (4.23) can be performed. This leads to $\delta$ functions what allows doing also the $\varepsilon$-integral. Furthermore, as shown by Migdal [6],

$$F_+(E,\varepsilon) \approx \left[1 - G\left(\frac{\varepsilon}{2\Delta}\right)\right] \delta(E-\mu) \qquad (4.27)$$

where (see eq. (3.5)),

$$\Delta = \Delta(\varepsilon_F)$$

and,

$$G(x) = \frac{\operatorname{arsinh}(x)}{x\sqrt{1+x^2}} \qquad (4.28)$$

Finally one obtains for the I-B part of the moment of inertia the following analytical expression [6,7],

$$\Theta_{I-B} = \Theta_{rigid} \left[1 - \frac{G_+ \omega_-^2 + G_- \omega_+^2}{\omega_-^2 + \omega_+^2}\right] \qquad (4.29)$$

where,

$$\omega_\pm = \omega_\perp \pm \omega_z \quad ; \quad G_\pm = G\left(\frac{\hbar \omega_\pm}{2\Delta}\right) \qquad (4.30)$$

and,

$$\Theta_{rigid} = \frac{\omega_z^2 + \omega_\perp^2}{\omega_z^3 \omega_\perp^4} \left(\frac{\mu^4}{12\hbar^3}\right) \qquad (4.31)$$

is the moment of inertia of rigid rotation. From (4.29) we see that

$$\lim_{\Delta \to 0} \Theta_{I-B} = \Theta_{rigid} \quad ; \quad \lim_{\Delta \to \infty} \Theta_{I-B} = 0.$$

The latter result is clearly unphysical and we will see how the account of the reaction of the pair field on the rotation will reestablish the physical situation.

**4.2 The Migdal term**

The density response corresponding to the Migdal term is obtained from (4.6),

$$(\rho_{1M})_{n,n'} = \frac{\xi_n \Delta_{1nn'} \Delta_{0n'} + \Delta_{0n} \Delta^*_{1nn'} \xi_{n'}}{2 E_n E_{n'} (E_n + E_{n'})} \tag{4.32}$$

In (4.32) we need to know $\Delta_1$ which we can gain from (4.7) in the following way; in the limit $\tau' \to \tau^+$ we obtain from $F_1^+$ the anomal density $\kappa_1^+$,

$$(\kappa_1^+)_{nn'} = -\frac{\xi_n H^*_{1nn'} \Delta_{0n'} + \Delta_{0n} H_{1nn'} \xi_{n'} + \Delta_{0n} \Delta_{1nn'} \Delta_{0n'} - (E_n E_{n'} + \xi_n \xi_{n'}) \Delta^*_{1nn'}}{2 E_n E_{n'} (E_n + E_{n'})} \tag{4.33}$$

In analogy with the non-rotating case where $\kappa_0 = \dfrac{\Delta}{2E}$, we also have,

$$(\kappa_1^+)_{nn'} = -\Delta^*_{1nn'} \left( \frac{1}{4E_n} + \frac{1}{4E'_{n'}} \right) \tag{4.34}$$

This relation stems from the fact that the quasiparticle energies contain the gap only in the form $\Delta \Delta^*$ and therefore there is no further first order correction, since in our case the external field is a time odd operator and thus,

$$\Delta_1^* = -\Delta_1 \equiv -i\Omega\chi \tag{4.35}$$

Equating (4.33) and (4.34) yields,

$$\frac{2\xi_n H^*_{1nn'}\Delta_{0n'} + 2\Delta_{0n}H_{1nn'}\xi_{n'} + 2\Delta_{0n}\Delta_{1nn'}\Delta_{0n'} + \left(\Delta^2_{0n} + \Delta^2_{0n'} + (\xi_n - \xi_{n'})^2\right)\Delta^*_{1nn'}}{2E_n E_{n'}(E_n + E_{n'})} = 0 \quad (4.36)$$

At this point we again exploit the fact that expression (4.36) is strongly peaked around the Fermi energy surface. Following [6], in analogy with (4.27), we have,

$$[E_n E_{n'}(E_n + E_{n'})]^{-1} \approx \frac{1}{\Delta^2} G\left(\frac{\varepsilon_n - \varepsilon_{n'}}{2\Delta}\right)\delta\left(\frac{\varepsilon_n + \varepsilon_{n'}}{2} - \mu\right) \quad (4.37)$$

With (4.35) we then obtain for (4.36),

$$\left[\frac{\langle n|\dot{L}_x|n'\rangle}{2\Delta} + \left(\frac{\varepsilon_n - \varepsilon_{n'}}{2\Delta}\right)^2 \chi_{nn'}\right]G\left(\frac{\varepsilon_n - \varepsilon_{n'}}{2\Delta}\right)\delta\left(\frac{\varepsilon_n + \varepsilon_{n'}}{2} - \mu\right) = 0 \quad (4.38)$$

where $\dot{L}_x$ stands for the time derivative of $L_x$. Summing on n and n' and following exactly the same line of semi-classical approximations as the ones used for the derivation of $\Theta_{I-B}$ one arrives at the following relation [7],

$$\int_{-\infty}^{+\infty} d\tau G(\tau) \int \frac{d^3p}{(2\pi\hbar)^3}\left[\frac{\dot{L}^{cl}_x}{2\Delta} - \frac{\ddot{\chi}(\tau)}{4\Delta^2}\right]\delta(\mu - H_{0cl}) = 0 \quad (4.39)$$

where $G(\tau)$ is the Fourier transform of $G(x)$ (4.28).

For the potential in (2.14) (4.39) is solved by,

$$\chi(\mathbf{R}) = \alpha R_y R_z \quad (4.40)$$

with,

$$\alpha = -2\Delta m \omega_+ \omega_- \frac{G_+ + G_-}{\omega^2_+ G_+ + \omega^2_- G_-} \quad (4.41)$$

Inserting this solution into (4.32) leads for the Migdal part of the moment of inertia to [6,7],

$$\Theta_M = \Theta_{rigid}\frac{\omega^2_+ \omega^2_-}{\omega^2_+ + \omega^2_-}\frac{(G_+ + G_-)^2}{\omega^2_+ G_+ + \omega^2_- G_-} \quad (4.42)$$

Together with (4.29) the expression for the moment of inertia is now complete. Let us again mention that we neglected the temperature dependence besides the one contained in $\Delta = \Delta(T)$, since all other T-dependence for $T < T_c$ is negligible. The moment of inertia can then be calculated as a function of deformation and temperature. For example it is immediately verified that for $\Delta \to \infty$ (4.42) yields the irrotational flow value,

$$\lim_{\Delta \to \infty} \Theta_M = \Theta_{irrot} = \Theta_{rigid}\left(\frac{\omega_\perp^2 - \omega_z^2}{\omega_\perp^2 + \omega_z^2}\right)^2 \qquad (4.43)$$

and therefore,

$$\lim_{\Delta \to \infty} \Theta = \lim_{\Delta \to \infty}\left(\Theta_{I-Bt} + \Theta_M\right) = \Theta_{irrot} \qquad (4.44)$$

what is the correct physical result.

## 5. Current distribution

One other quantity, which may be interesting also from the experimental point of view, are the current distributions of the superfluid rotating gas. Indeed after a sudden switch off of the (rotating) trap the atoms will expand keeping memory of their rotational state. So if the velocity distribution of the expanding atoms can be measured, one may be able to deduce the rotational motion the atoms have had before the trap was taken away. The current distribution, as we will see, depends, as the moment of inertia, strongly on the superfluid state of the gas. In order to calculate the current distribution we first write down the Wigner function of the density response which can easily be red off from the formulas given in Section 4. In obvious notations we obtain [7],

$$\rho_{1I-B}(\mathbf{R},\mathbf{p}) = \Omega[\mathbf{R}\times\mathbf{p}]_x \delta(\mu - H_{0cl}) \qquad (5.1)$$

$$-\Omega\left[\frac{\omega_+ G_- - \omega_- G_+}{2\omega_z} R_y p_z - \frac{\omega_+ G_- + \omega_- G_+}{2\omega_\perp} R_z p_y\right]\delta(\mu - H_{0cl})$$

$$\rho_{1M}(\mathbf{R},\mathbf{p})=\frac{\hbar^2}{2m}\frac{\alpha}{\Delta}\Omega\left[\frac{\omega_+G_+-\omega_-G_-}{\omega_z}R_y p_z+\frac{\omega_+G_++\omega_-G_-}{\omega_\perp}R_z p_y\right]\delta(\mu-H_{0cl}) \qquad (5.2)$$

With the usual definition of the current,

$$\mathbf{j}(\mathbf{R})=\int\frac{d^3p}{(2\pi\hbar)^3}\frac{\mathbf{p}}{m}\rho(\mathbf{R},\mathbf{p}) \qquad (5.3)$$

one obtains,

$$j_y^{I-B}=-\rho_{TF}(\mathbf{R})R_z\Omega\left[1-\frac{\omega_+G_-+\omega_-G_+}{2\omega_\perp}\right] \qquad (5.4a)$$

$$j_z^{I-B}=-\rho_{TF}(\mathbf{R})R_y\Omega\left[1-\frac{\omega_+G_--\omega_-G_+}{2\omega_z}\right] \qquad (5.4b)$$

$$j_y^M=-\rho_{TF}(\mathbf{R})R_z\Omega\left[\frac{\omega_+\omega_-(G_-+G_+)}{\omega_+^2 G_++\omega_-^2 G_-}\frac{\omega_+G_-+\omega_-G_+}{\omega_\perp}\right] \qquad (5.5a)$$

$$j_z^M=-\rho_{TF}(\mathbf{R})R_y\Omega\left[\frac{\omega_+\omega_-(G_-+G_+)}{\omega_+^2 G_++\omega_-^2 G_-}\frac{\omega_+G_--\omega_-G_+}{\omega_z}\right] \qquad (5.5b)$$

with of course, $j_x=0$.

Again we see that in the limit $\Delta\to\infty$ the current approaches the correct irrotational flow limit,

$$\mathbf{j}\underset{\Delta\to\infty}{\longrightarrow}-2\rho_{TF}\Omega\frac{\omega_\perp^2-\omega_z^2}{\omega_\perp^2+\omega_z^2}\nabla(r_y r_z) \qquad (5.6)$$

whereas in the limit of $\Delta\to 0$ we obtain a rigid body current. As we have seen for $\Theta$, as a function of temperature and deformation, we easily can go from one limit to the other. We therefore show in Fig. 4 a,b the current distribution for the two extreme cases of irrotational and rigid body flow in the laboratory frame. We see that the flow pattern is completely different in the two cases. As a function of temperature one continuously can pass from one flow pattern to the other. The point we want to make is that for small deformations, as can be seen from (5.6) there is almost no irrotational current for low temperatures and this should be detectable experimentally.

6. **Results**

In Fig. 5 we show $\Theta$ as a function of $\Delta(T)$ (and with Fig. 3 as a function of T) for several values of the deformation parameter $\delta^{-1}$. We see that, not unexpectedly, for $T \approx 0$ the system is in the irrotational flow limit. This means that the number of levels in a range $\Delta(0)$ is very high and thus our statistical assumptions well verified. Only for temperatures very close to the critical temperature where $\Delta \leq 10\,\text{nK}$ there is then a strong turn over to the rigid moment of inertia. We also see that the details depend in an appreciable way on the deformation: the greater $\delta$ the earlier starts the bending to the rigid value. Tracing the moment of inertia as a function of deformation for different values of the gap yields the curves shown in Fig. 5. We see that, independently of the gap value, for sufficiently strong deformation the moment of inertia will always end up in its rigid body value. This is easily understandable since in the limit of very strong prolate deformation the system becomes essentially one dimensional. Turning a fluid contained in a rod shaped trap, whether superfluid or not, one always has to turn the whole body of the fluid equivalent to rigid body rotation. We therefore see that the moment of inertia of a trapped superfluid gas of atomic fermions can span a very wide range of values going from almost zero at low temperatures and low deformations of the trap to the rigid body value at temperatures close to $T_c$ and/or at strong deformations. Once the problem of putting the trapped gas into rotation is mastered experimentally and the question of measuring the moment of inertia solved one will then be able to demonstrate very clearly in changing the system parameters whether or not the system is in a superfluid state.

7. **Discussion and conclusions**

In this work we transcribed work of nuclear physics for the evaluation of the moment of inertia of a rotating superfluid nucleus to the situation of a rotating gas of trapped superfluid atomic fermions. The question of how to experimentally detect the superfluid state of such a gas has recently been discussed in the literature to quite some extent. We argue that, like for nuclei, one unambiguous indication for superfluidity can come from the study of the moment of inertia as a function of such system parameters as temperature and trap deformation. We have seen that the value of $\Theta$ can run through large variations as the gap, as a function of T, varies from zero to its maximal value. The same is true for the variation of $\Theta$ with the deformation of the trap. The direct determination of the moment of inertia may be difficult. However the variation at constant deformation of the rotational energy,

$$E_{rot} = \frac{\Theta}{2}\Omega^2 \qquad (7.1)$$

as a function of the temperature, i.e. as a function of $\Delta(T)$ should be measurable scanning the range from $\Theta_{rigid} > \Theta > \Theta_{irrot}$. Indeed taking for example a rotational frequency of $\Omega = 1.6\,s^{-1}$ (which should be compared to our harmonic oscillator frequency $\omega_0 \approx 7\,s^{-1}$), we obtain for $\delta = \omega_z/\omega_\perp = 0.5176$ the following variation (see Fig. 6) for,

$$\varepsilon = \frac{E_{rot}}{\frac{1}{2}m\omega_0^2\langle R^2 \rangle} \qquad (7.2)$$

as a function of $\Delta(T)$.

We see that the rotational energy makes about 30% of the harmonic oscillator energy for rigid rotation, dropping by a factor of ~3 for superfluid rotation. Supposing an experimental error of ~10% in the measurement of the rotational energy obtainable from a sudden switch off of the trap, one deduces that the variation as a function of T of the rotational energy should be in the range of experimental accuracy. However, other ways to detect the rotational dynamics may be imagined. Suppose an almost spherical rotating trap is suddenly stopped from rotating. For temperatures T>$T_c$ there is rigid rotation and the Fermi gas will continue rotating for a while. On the other hand for T<<$T_c$ the trap will rotate but the superfluid gas will not be in a rotational state neither before nor after the stop of the trap. Whether the gas rotates or not may eventually be detectable by optical means.

In above discussion we ignored the possibility of vortex formation. The onset of instability versus vortex formation in a finite Fermi system is not a completely easy task and we will postpone such an investigation to a future work. However, since the rotational frequencies $\Omega$ considered in the present work are much smaller than the oscillator constant $\omega_0$ ($\Omega/\omega_0 < 0.2$) we may hope that our results will not be spoiled by the appearance of vortices. An indication can also come from the case of trapped Bosons where vortices, depending somewhat on the number of atoms, do not appear for values $\Omega/\omega_0 < 0.5$ (see ref.[2]).

From the above discussion we see that it may well be in experimental reach to reveal an eventual superfluid state of the gas once the technique of putting the trap into rotation will be mastered experimentally.

In short we proposed in this work to measure the dynamics of a rotating trapped gas of atomic Fermions as a function of temperature and deformation to detect whether the system is in a superfluid state or not. Quite detailed and quantitative calculations for the moment of inertia and velocity distributions have been presented. Other quantities well studied in the case of rotating superfluid nuclei [5] such as Yrast lines, even-odd effects, particle alignment, etc., may also become of interest in this case.

Acknowledgements: We gratefully acknowledge very useful discussions with C. Gignoux , S. Stringari and W. Zwerger and a careful reading of the manuscript by M. Durand. One of us (X.V) also acknowledges financial support from DGCYT (Spain) under grant PB95-1249 and from the DGR (Catalonia) under grant GR94-1022.

**Figure captions**

**Figure 1 :** Density profiles, for the case of a spherical trap, of the non-interacting case (full line), the interacting case once calculated exactly from (2.3) (crosses) with $V_{ex}$ given by (2.10) and once using the variationally determined harmonic oscillator potential(open squares). Squeezing the variational $\omega$ by 6% yields a density which lies on top of the exact TF solution.

**Figure 2 :** The gap for a spherical trap as a function of the radius

**Figure 3 :** Ratio of the energy gap to the gap at T=0°K as a function of temperature

**Figure 4 a and b :** The current distribution for the two extreme cases of irrotational and rigid body flow in the laboratory frame (note the change of scale between the two figures)

**Figure 5 a and b :** The moment of inertia as a function of deformation for different values of the gap (a) and as a function of the gap for different values of the deformation (b)

**Figure 6** : The variation of $\varepsilon$ (eq. 7.2), ratio of the rotational energy of the gas and the harmonic oscillator energy, as a function of the gap energy $\Delta(T)$ for different values of deformation $\delta = \dfrac{\omega_z}{\omega_\perp}$ and rotational frequency $\Omega$.

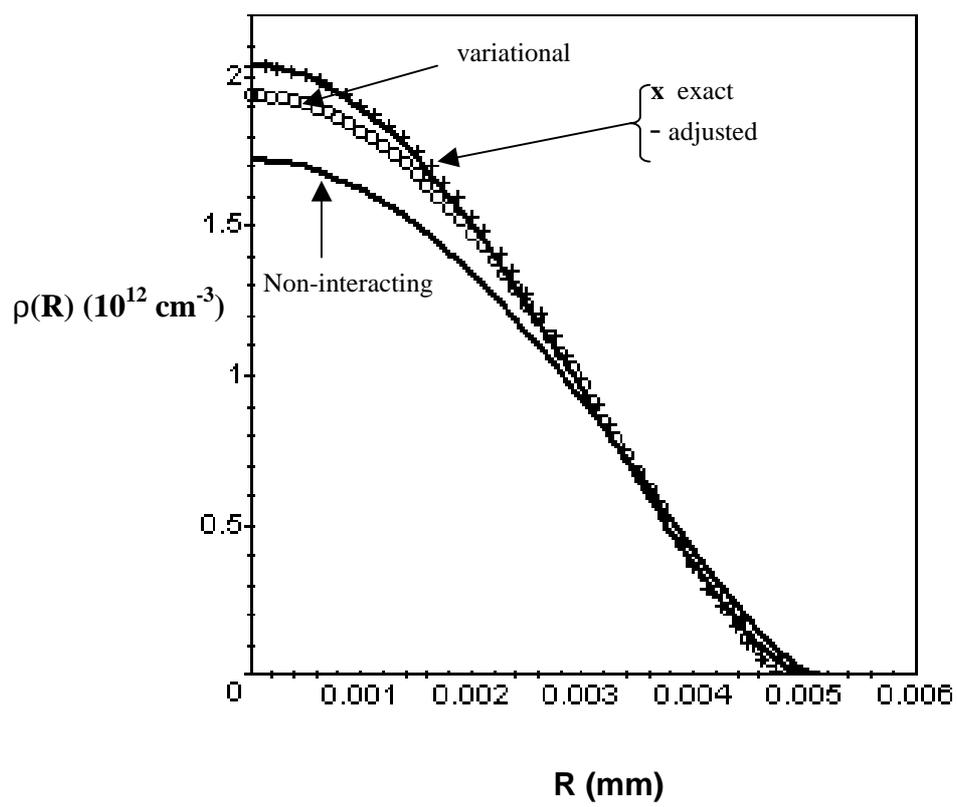

**Figure 1**

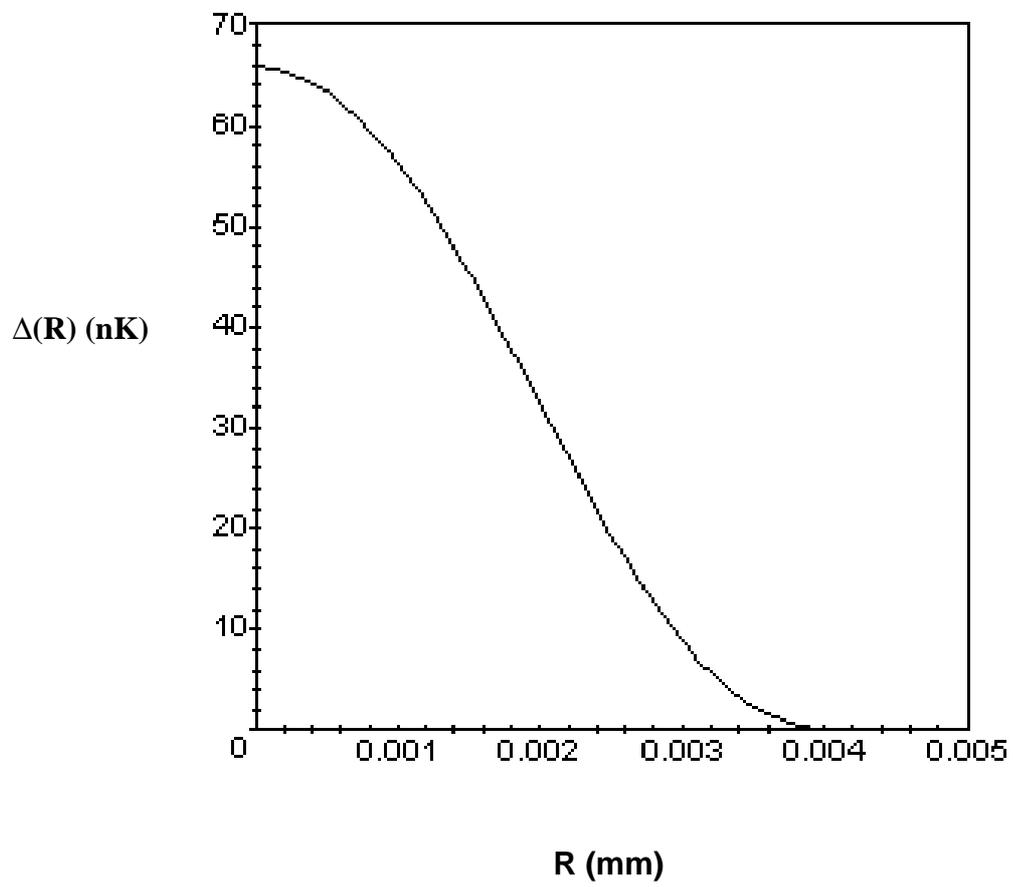

**Figure 2**

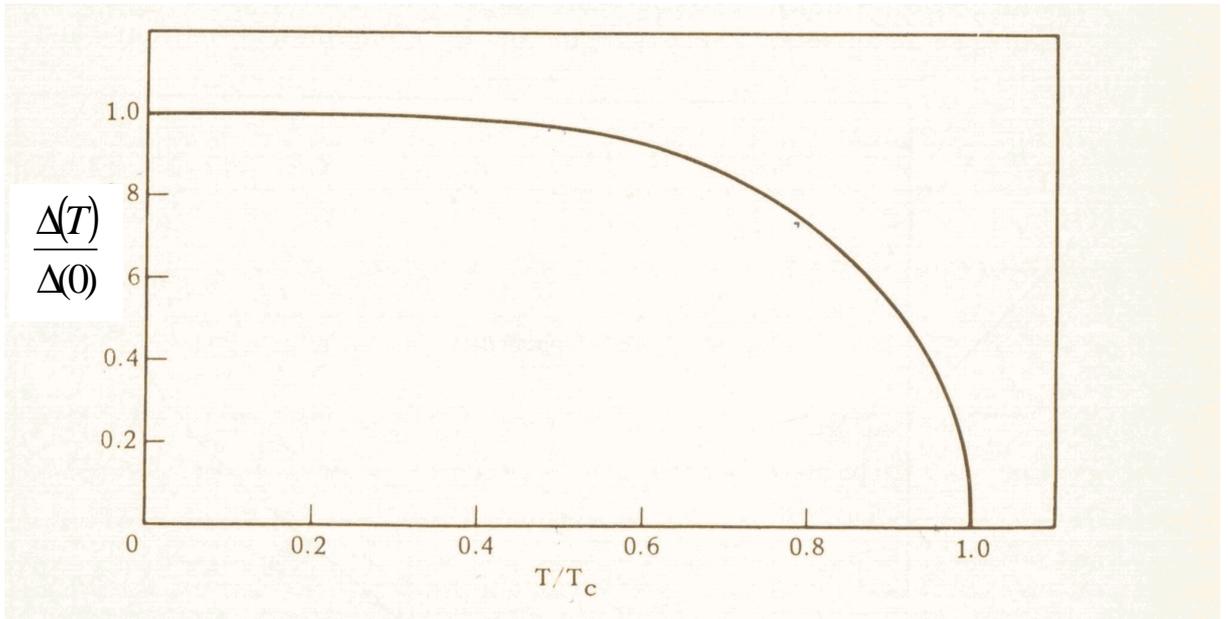

**Figure 3**

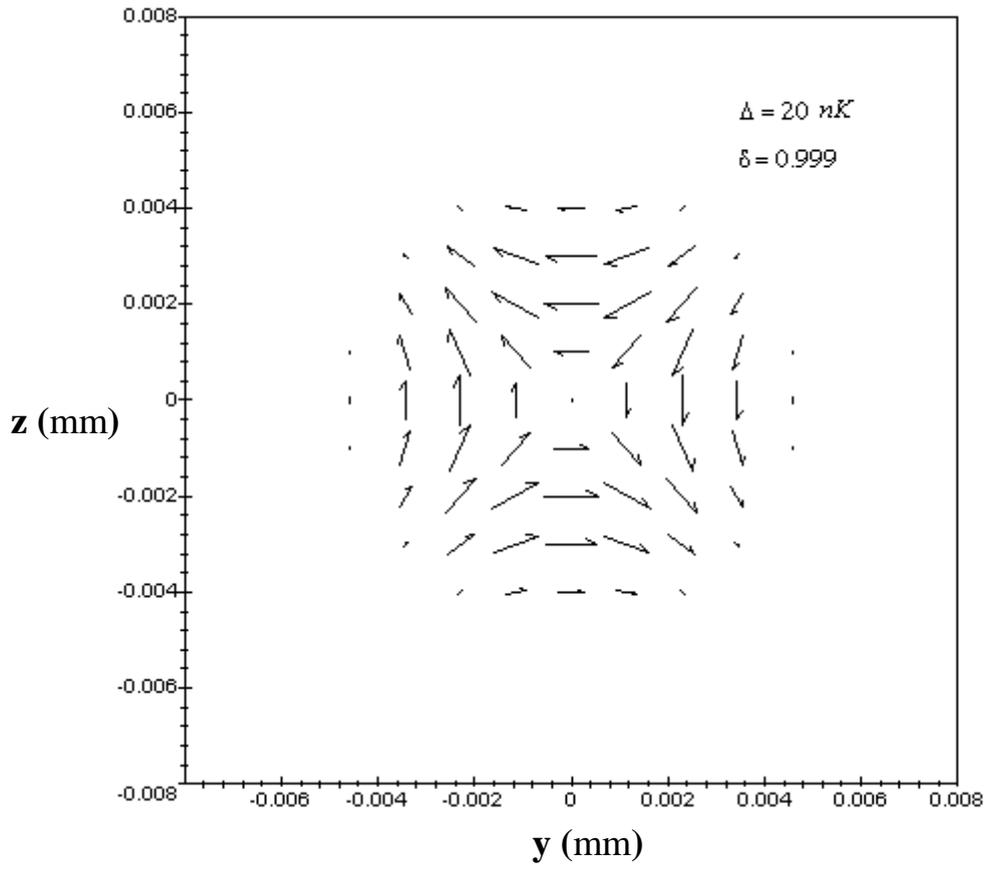

z (mm)

y (mm)

**Fig 4 a and b**

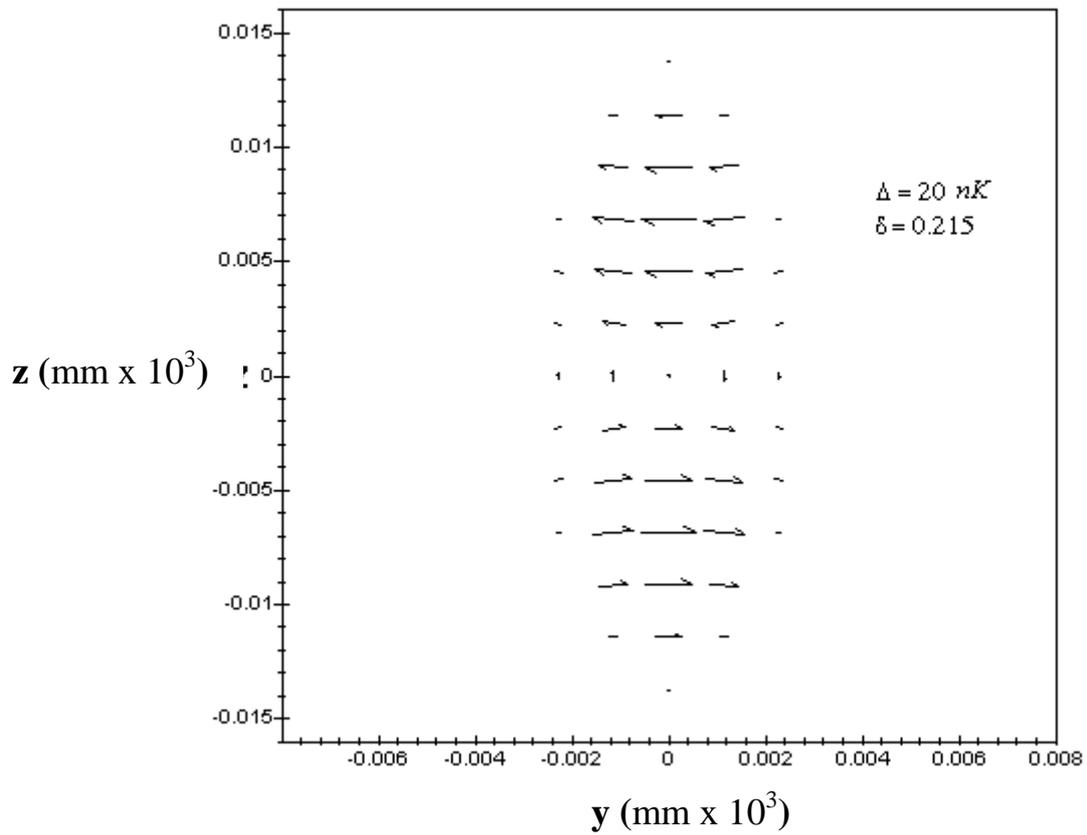

z (mm x 10³)

y (mm x 10³)

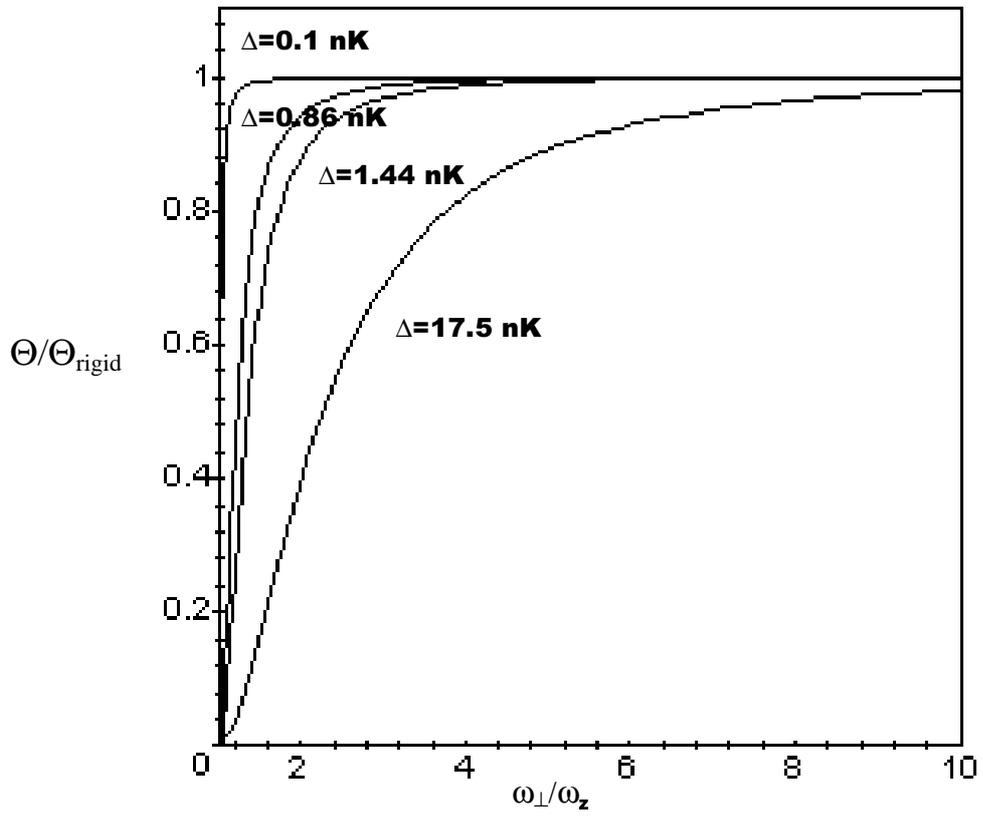

**Figure 5 a and b**

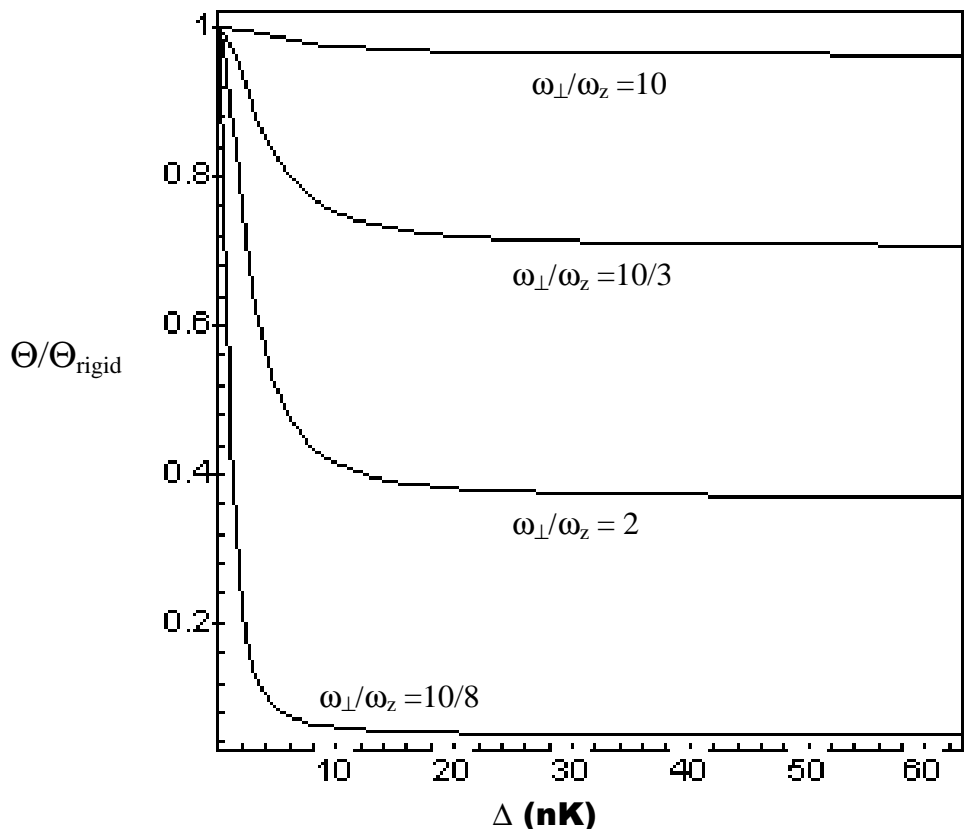

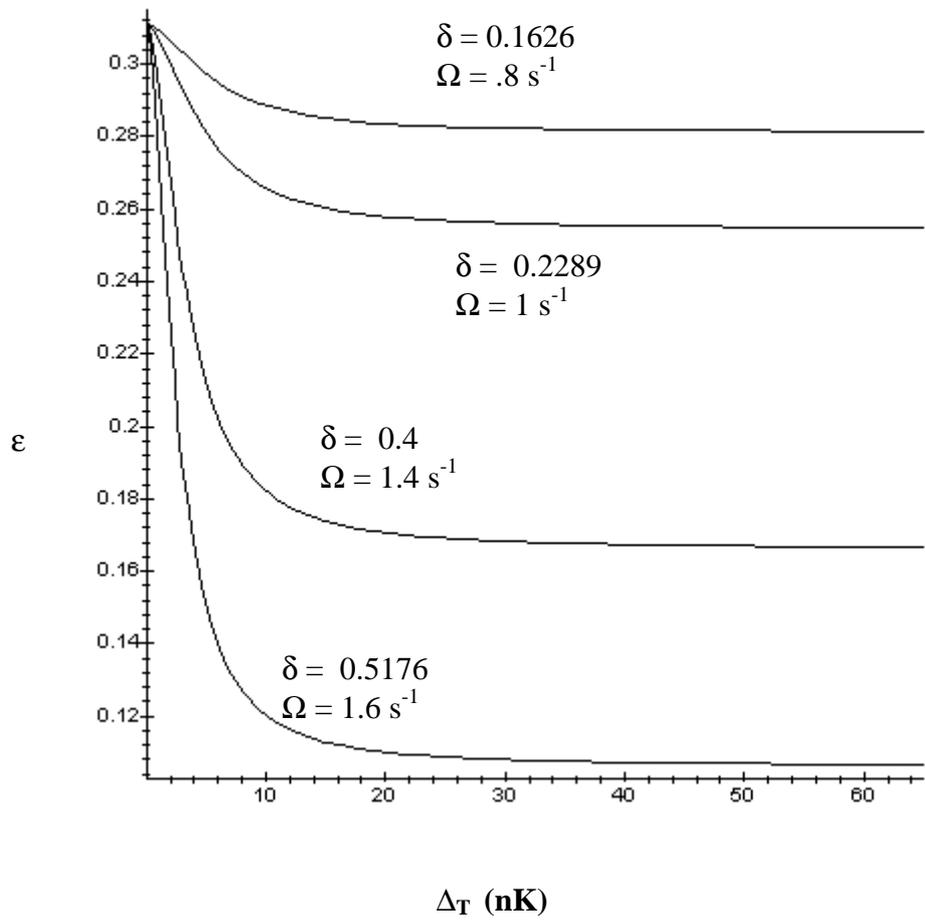

**Figure 6**